\documentclass[aps,prb,preprint,tightenlines,showpacs]{revtex4}
\makeatother

\usepackage{amsmath}
\usepackage{graphicx}
\usepackage{dcolumn}
\usepackage{bm}

\begin{document}

\title{Quantum Monte Carlo study of small pure and mixed spin-polarized 
tritium clusters}

\author{I. Be\v{s}li{\'c}}
\affiliation{Faculty of Natural Sciences, University of Split, HR-21000 Split,
Croatia}
\author{L. Vranje\v{s} Marki{\'c}}
\affiliation{Faculty of Natural Sciences, University of Split, HR-21000 Split,
Croatia}
\affiliation{ Institut f\"ur Theoretische Physik, 
Johannes Kepler Universit\"at, A-4040 Linz, Austria}
\author{J. Boronat}
\affiliation{Departament de F\'\i sica i Enginyeria Nuclear, Campus Nord
B4-B5, Universitat Polit\`ecnica de Catalunya, E-08034 Barcelona, Spain}
\date{\today}

\begin{abstract}
We have investigated the stability limits of small spin-polarized 
clusters consisting of up to ten spin-polarized tritium T$\downarrow$ atoms
and the mixtures of T$\downarrow$ with spin-polarized deuterium
D$\downarrow$ and  hydrogen H$\downarrow$ atoms. All of our calculations
have been performed using the variational and diffusion Monte Carlo methods.
For clusters with D$\downarrow$ atoms, the released node procedure is used
in cases where the wave function has nodes. In addition to the energy, we
have also calculated the structure of small clusters using unbiased
estimators. Results obtained for pure T$\downarrow$ clusters are in good
accordance with previous calculations, confirming that the trimer is the
smallest spin-polarized tritium cluster. Our results show that  mixed
T$\downarrow$-H$\downarrow$ clusters having up to ten atoms  are unstable
and that it takes at least three tritium atoms to bind one, two or three
D$\downarrow$ atoms. Among all the considered clusters, we have found no
other Borromean states except the ground state of the
T$\downarrow$ trimer.
\end{abstract}


\maketitle

\section{Introduction}

Electron spin-polarized hydrogen (H$\downarrow$) has extreme quantum nature
due to its very small mass and weakly attractive potential. It remains in
gas phase even in the limit of zero temperature and up to pressures of
about 170 bar,~\cite{hydrogen} after which it solidifies. Stwalley and
Nosanow~\cite{StwaleyNosanow} suggested in 1976 the use of H$\downarrow$
for achieving a Bose-Einstein condensate (BEC) state, which was finally achieved
for hydrogen in 1998 by Fried \textit{et al.}.~\cite{fried} Heavier
isotopes of hydrogen,  spin-polarized deuterium (D$\downarrow$) and tritium
(T$\downarrow$) also show a remarkable   quantum behavior. D$\downarrow$
atoms, which obey Fermi statistics, have nuclear spin one and thus the
zero-pressure state of  bulk D$\downarrow$ depends on the number of
occupied nuclear spin states. Previous theoretical
estimations~\cite{panoff,flynn,skjetne} have shown that  (D$\downarrow_1$)
with only one occupied nuclear spin state is gas at zero pressure, while
bulk D$\downarrow$ with two (D$\downarrow_2$)  and three 
(D$\downarrow_3$)  equally occupied nuclear spin states  remains liquid at
zero pressure and zero temperature.  Spin-polarized tritium, which obeys
Bose statistics, is expected to be  liquid~\cite{etters,miller,tritium} due
to its larger mass. 

Recently, Blume \textit{et al.}~\cite{blume} have studied microscopic
properties of tritium  (T$\downarrow$)$_N$ clusters with up to $N$=40 atoms 
using the diffusion Monte Carlo (DMC) method and compared them to bosonic
$^4$He$_N$ clusters. In that work, it is shown that tritium clusters
(T$\downarrow$)$_N$ are even more weakly bound than helium clusters
$^4$He$_N$ with the same number of atoms. Furthermore, their results
suggest the use of   T$\downarrow$ as a new BEC gas with the advantage of a
nearly exact  knowledge of its interatomic potential and the possibility of
manipulating the strength of the interactions 
via Feshbach resonances. In addition, it has been shown~\cite{blume}
that the trimer (T$\downarrow$)$_3$ is the smallest spin-polarized tritium
cluster with a ground-state energy of only $-4.2(7)$ mK. It is thus an
example of Borromean or halo state,~\cite{fedorov,jensen} because the
two-body T$\downarrow$ system does not have a bound state. The prediction
of a total zero angular momentum bound state of the spin-polarized atomic
tritium trimer has been recently confirmed by Salci 
\textit{et al.}~\cite{salci} using the finite element method. 
Importantly, the use of this accurate few-body approach leads to conclude
that there are not any other bound or shape resonant states with zero and 
nonzero angular momentum.~\cite{salci}

Unlike the case of pure $^3$He, $^4$He and mixed $^3$He-$^4$He clusters,
which have been the subject of both experimental and theoretical 
study,~\cite{expmix,review,fantoni,mixhel,ester} nothing is known about the
stability of mixed or other pure spin-polarized hydrogen clusters different
from tritium ones. $^3$He has almost the same mass as T$\downarrow$, but
because of its small mass and Fermi statistics it takes at least 30 $^3$He
atoms to form a bound state.~\cite{ester} Recently, Hanna and
Blume~\cite{hanna} have studied the near-threshold behavior of weekly bound
three-dimensional bosonic clusters with up to 40 atoms interacting
additively through two-body van der Waals potentials modeled after the
T$\downarrow$-T$\downarrow$ interaction. Among other results, they found a
four-parameter fit for the critical mass needed to bind a bosonic cluster
of $N$ atoms.  From this, it follows that if D$\downarrow$ were a boson,
due to its smaller mass compared to T$\downarrow$, it would take at least
13 atoms to form a bound cluster (D$\downarrow)_N$.  Considering  that
D$\downarrow$ obeys Fermi statistics we expect that, similarly to the case
of $^3$He,~\cite{ester} it will take more than 30 D$\downarrow$ atoms  to
form a self-bound cluster.   On the other hand, although H$\downarrow$
obeys Bose statistics, its approximately three times larger zero-point
energy (with respect to  T$\downarrow$)  implies that clusters of
H$\downarrow$ are not  stable.~\cite{hanna} This statement is in agreement
with the gas nature of bulk H$\downarrow$ at zero
temperature.~\cite{hydrogen} Therefore, hydrogen-tritium and
deuterium-tritium clusters are reasonable choices for getting stable
self-bound spin-polarized clusters.

If tritium could be experimentally prepared to achieve a 
BEC state, as suggested in Ref. \onlinecite{blume}, the presence of
a finite fraction of deuterium would be a new example of a Bose-Fermi
mixture with BEC. In fact, Bose-Fermi mixtures composed with alkalines are
nowadays studied in several laboratories and the main physics underlying
them is well known.~\cite{stefano_rev} Tritium-deuterium mixtures would be
even more interesting since they are isotopic and with a very well known
interaction. Dilute vapors are also candidates for searching the elusive Efimov
states,~\cite{efimov} and recently experimental evidence of its existence
has been observed in an ultracold gas of cesium atoms.~\cite{cesium} In
the case of pure tritium, it has been proved that such state does not
exist~\cite{blume} but mixed isotopic hydrogen systems could offer new
possibilities for continuing the search of trimers with Efimov character.

In this article, we report energetic and structural properties of small
spin-polarized tritium-deuterium and tritium-hydrogen clusters, obtained
using the diffusion Monte Carlo method. It is our main goal to determine which
of these clusters are stable. As many of them extend beyond the classically
forbidden region they can be considered as quantum halo systems. We are
particularly interested in finding  among these halo clusters mixed
Borromean or  "super-Borromean''~\cite{hanna} systems for which all
subsystems are unbound.  So far, $^3$He$_2$K is the only mixed molecular
system for which the Borromean state has been predicted theoretically.~\cite{li}

In Sec. II, we briefly describe the DMC method and the trial wave functions
used for importance sampling. Due to the fermionic nature of D$\downarrow$ ,
the fixed-node and released-node methods are used in cases where the 
wave function has
nodes. Sec. III reports the results obtained, and finally,  
Sec. IV comprises the summary of the results and main conclusions.


\section{Method}

We consider that all the H$\downarrow$, D$\downarrow$, and T$\downarrow$
atoms  interact with the spin-independent central triplet pair potential
$b$$^3\Sigma_u^+$. This interaction  has been determined in an essentially 
exact way by Kolos and Wolniewicz,~\cite{kolos} and recently extended to
larger interparticle distances by Jamieson \textit{et al.}
(JDW).~\cite{jamieson} We have used a cubic spline interpolation between
JDW data and smoothly connected the points to the long-range behavior as
calculated by Yan \textit{et al.}.~\cite{yan} The same form has been used
in the recent DMC calculation of bulk H$\downarrow$,\cite{hydrogen} work in
which  a comparison  between different potentials
employed in the past is also  reported. On the other hand, we have checked that the  addition
of mass-dependent adiabatic corrections (as calculated by
Kolos and Rychlewski~\cite{kolos2}) to the JDW potential does not modify
the energy of spin-polarized hydrogen clusters.

The starting point of the DMC method is the Schr\"odinger equation 
written in imaginary time,
\begin{equation}
-\hbar \frac{\partial \Psi(\bm{R},t)}{\partial t} = (H- E) \Psi(\bm{R},t) \ ,
\label{srodin}
\end{equation}
where  $E$ is a constant acting as a reference energy,
$\bm{R} \equiv (\bm{r}_1,\ldots,\bm{r}_N)$ is a \textit{walker} in Monte
Carlo terminology, and $H$ is the $N$-particle Hamiltonian
\begin{equation}
 H = -\sum_{i=1}^{N}\frac{\hbar^2}{2m_i}  \bm{\nabla}_i^2 + 
 \sum_{i<j}^{N} V(r_{ij})  \ .
\label{hamilto}
\end{equation}

As usual in the method, a trial wave function $\psi(\bm{R})$  is 
introduced for importance sampling and then the Schr\"odinger equation
(\ref{srodin}) is rewritten in terms of $\Phi(\bm{R},t)= \Psi(\bm{R},t)
\psi(\bm{R})$. The resulting equation is solved  stochastically by
considering a short-time approximation for the Green's function.  In the
limit $t\rightarrow\infty$ (for long simulation times), only the lowest
energy eigenfunction not orthogonal to $\psi(\mathbf{R})$ survives and
then the sampling of the ground state is effectively achieved.

The trial wave function used for the simulation of the pure T$\downarrow$
clusters is of Jastrow form, $\psi_{\text J}(\bm{R}) = \prod_{i<j}^{N}
f_T(r_{ij})$,  where $f_T(r_{ij})$ describes correlations between pairs of
tritium atoms. Similarly,  wave functions for the study of mixed
tritium-hydrogen clusters, containing  $N_1$ H$\downarrow$ atoms and $N_2$
T$\downarrow$ atoms ($N_1+N_2=N$), are constructed as  a product of 
two-body correlation functions between all the pairs, 
\begin{equation}
\psi(\mathbf{R};N_1,N_2)=\prod_{\stackrel{{\scriptstyle
i,j=1}}{i<j}}^{N_1}f_H(r_{ij})\prod_{\stackrel{{\scriptstyle i,j=N_1+1}}{i<j}}^{N}f_{T}(r_{ij})
\prod_{i=1}^{N_1}\prod_{j=N_1+1}^{N}f_{HT}(r_{ij}) \ ,
\end{equation}
where $f_{H}(r)$ describes the two-body correlations between H$\downarrow$
atoms and  $f_{HT}(r)$ accounts for the H$\downarrow$-T$\downarrow$ pairs.
In case of mixed  (D$\downarrow$)$_{N_1}$(T$\downarrow$)$_{N_2}$ clusters,
the total wave-function  is constructed as a product of an antisymmetric
function corresponding to deuterium ($\psi_A$) and Jastrow factors,
\begin{equation}
\psi(\mathbf{R};N_1,N_2)=\psi_A 
\prod_{\stackrel{{\scriptstyle i,j=1}}{i<j}}^{N_1}f_D(r_{ij})
\prod_{\stackrel{{\scriptstyle i,j=N_1+1}}{i<j}}^{N}
f_{T}(r_{ij})\prod_{i=1}^{N_1}\prod_{j=N_1+1}^{N}f_{DT}(r_{ij}) \ ,
\end{equation}
where $f_{D}(r)$ describes two-body correlations between D$\downarrow$
atoms and $f_{DT}(r)$ two-body correlations  for the
D$\downarrow$-T$\downarrow$ pairs. For the function $\psi_A$ it is enough
to consider the product of  Slater determinants for each one of the nuclear
spin states. Consequently, $\psi_A$ has different form in the case of 
D$\downarrow_2$ and D$\downarrow_3$ mixed D$\downarrow$-T$\downarrow$
clusters. For example, in the  case of
(D$\downarrow_2)_{N_1}$(T$\downarrow)_{N_2}$ clusters, $\psi_A=1$ for
$N_1=1,2$, $\psi_A$=$x_1-x_2$  for $N_1=3$, and
$\psi_A$=$(\bf{r}_1-\bf{r}_2)(\bf{r}_3-\bf{r}_4)$ for $N_1$=4, while for  
(D$\downarrow_3)_{N_1}$($T\downarrow)_{N_2}$ clusters $\psi_A=1$ for
$N_1=1,2,3$, $\psi_A=x_1-x_2$ for $N_1$=4 and
$\psi_A$=$(\bf{r}_1-\bf{r}_2)(\bf{r}_3-\bf{r}_4)$ for $N_1$=5.   

We have worked with three different models for the two-body correlation 
function $f_T(r)$, 
\begin{equation}
f_T(r)=\exp [-b_1 \exp(-b_2r) -b_3r] \ ,
\label{trial1}
\end{equation}
\begin{equation}
f_T(r)=\frac{1}{r} \,
\exp\left[-\left(\frac{\alpha}{r}\right)^{\gamma}-sr\right] \ ,
\label{trial2}
\end{equation}
\begin{equation}
f_T(r)=\exp\left[-\left(\frac{b}{r}\right)^{5}-sr\right] \ ,
\label{trial3}
\end{equation}
where $b_1$, $b_2$, $b_3$, $\alpha$, $\gamma$, $s$ and $b$ are variational
parameters. They have been obtained  by optimizing the variational energy
calculated with the VMC method. Finally, we found the  form (\ref{trial2})
to be optimal for clusters from three to five T$\downarrow$ atoms,  while
the other two (\ref{trial1},\ref{trial3}) suit better for larger clusters. 
The same function, with
different variational  parameters, has then been used for all the other 
two-body correlation functions ($f_H(r)$, $f_D(r)$, $f_{HT}(r)$,
$f_{DT}(r)$). 
For example, in case of pure T$\downarrow$ clusters $(\alpha, \gamma, s)$ range from (3.9,4.0,0.0012) for trimer to (4.0,3.7,0.001) for (T$\downarrow)_5$ and $(b,s)$ range from (3.55,0.065) for (T$\downarrow)_6$ to (3.6,0.05) for (T$\downarrow)_{10}$. The VMC recovers from 70\% to 90\% of the DMC energy, except in case of (T$\downarrow)_3$ where the VMC energy is only -0.85(8) mK. In case of mixed clusters the parameters in the $f_T(r)$ function do not change significantly, so for example in case of the form (\ref{trial3}) $b$ assumes values from 3.5 to 3.62 and $s$ from 0.07 to 0.05, going from smaller to larger clusters, respectively. At the same time, in the function $f_{TH}(r)$,  $b$ goes from 3.58 to 3.7 and $s$ is around 0.003. Similar behavior of variational parameter $b$ is obtained for the other correlation functions, while $s$ is around 0.01 in $f_{DT}(r)$ and around 0.003 in $f_{D}(r)$. In most cases of mixed clusters VMC recovers from 60\% to 90\% of the DMC energy, exceptions being smaller clusters with more than two D$\downarrow$ atoms and the clusters having only three T$\downarrow$ atoms, where the VMC energies are further from the DMC ones.

The DMC method we have used is accurate to second order in the time step
$\Delta t$.~\cite{boro}  The ground-state energies have been calculated for
several time-steps and  then  extrapolated  to zero $\Delta t$ to remove
any possible time-step bias. We have also studied the optimal mean walker
population and finally chosen $N_w=1000$ in order to eliminate any bias
coming from it. 

Apart from statistical uncertainties, the energy of the bosonic clusters is
exactly calculated.   In the case of mixed clusters with more than three
D$\downarrow_3$ or two D$\downarrow_2$ atoms, the sign problem appears
because  the trial wave function changes sign. Since the Monte Carlo method
requires that  $f(\mathbf{R},t)=\Psi(\mathbf{R},t) \psi(\mathbf{R})\geq 0$,
we have first used the so-called  fixed-node approximation which allows
only the moves in which  $\psi$ and $\Psi$ change sign together, thus
fixing the nodes. In this way, an upper bound to the energy is
obtained.~\cite{boundfn} In a subsequent step, we  have removed the nodal 
constraint imposed by the trial wave function by using the released-node
method. We have adopted the methodology used previously in liquid helium 
calculations,~\cite{boro2} and recently employed in the study of  small
mixed helium clusters.~\cite{mixhel} In this approach, walkers are allowed
to cross the nodes imposed by the trial wave function and survive for a
finite lifetime. In order to achieve an effective crossing of the nodal
surface an auxiliary guiding  wave function $\psi_{g}$ is introduced in the
DMC calculation. This function is positive  everywhere, different from zero
in the nodes, and approaches $|\psi|$ away from the nodes. Like in Refs.
\onlinecite{boro2,mixhel}, we have taken
$\psi_{g}=\psi_{J}(\bf{R})(\psi_{A}^2(\bf{R})+ \textit{a}^2)^{1/2}$, which
satisfies the above condition for suitable choices of $a$.  The
released-node energy is estimated through an exponential fit $E(t)=E_{r} +
A e^{-(t/\tau)}$ to the DMC data, with $t$ the released time.  In all cases
where the RN method is used in this work the difference between  the last
calculated point in released time and $E_{r}$ is of the same order as the
statistical noise. The behavior of $a$ and the maximum released time needed
to achieve the asymptotic limit of the released energy has been similar to
the case studied in  Ref. \onlinecite{mixhel}. Also, like in Ref.
\onlinecite{mixhel}, we find the RN energies to be very close to the FN
ones. In addition, in the case of $(T\downarrow)_4(D\downarrow_2)_3$ we
have verified that the inclusion of backflow correlations in  $\psi_A$ does
not change the final FN results.   

\section{Results}

\begin{figure}[p]
\centering
        \includegraphics[width=8.5cm]{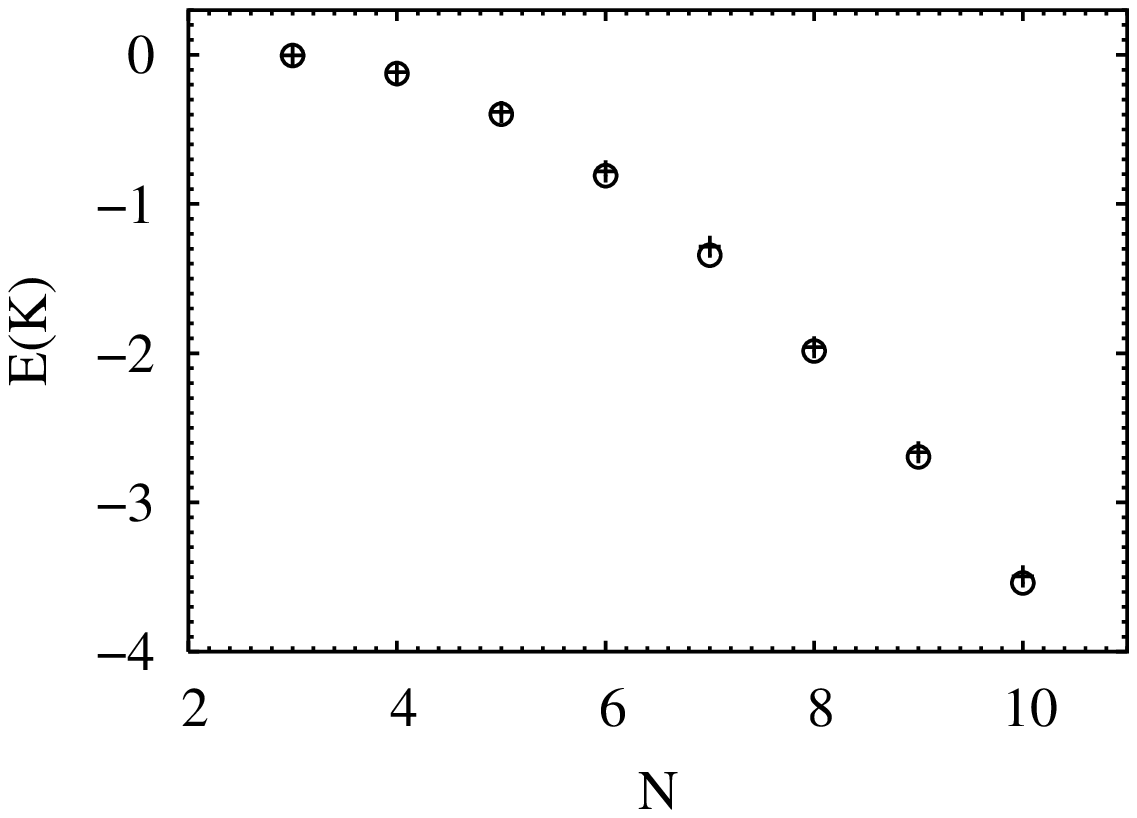}
        \caption{I. Be\v{s}li{\'c} {\it et al.} }
\label{fig:blume}	
\end{figure}

DMC results for the ground state energy of pure and mixed
T$\downarrow$-H$\downarrow$ clusters are given in Table
\ref{tab:energiesTH}. Our results for pure tritium T$\downarrow$ clusters
are compared to the results of Blume \textit{ et. al.}~\cite{blume} in Fig.
\ref{fig:blume}. We obtain a good agreement with these published data, 
although our results are slightly lower for all $N$.  Underlying this
difference is the fact that in Ref. \onlinecite{blume} a damped three-body
Axilrod-Teller potential term~\cite{ATterm} is  introduced in the
Hamiltonian, raising the trimer energy 1.6\% and the energy of the cluster
with 40 particles 6\%. We also find that the trimer is the smallest
spin-polarized tritium cluster with an energy of only $-4.8(2)$ mK,
confirming it to be a halo state. In order to test the sensitivity of the
present results  to the details in the interaction potential, we have also
calculated the energy with the potential  that Silvera and
Goldman~\cite{silvera} constructed as a fit to the older Kolos and
Wolniewicz data.~\cite{kolos} It is worth mentioning that a  full
comparison between the different potentials used in the literature for
studying bulk H$\downarrow$  can be found in the Ref. \onlinecite{hydrogen}.
Using the Silvera model, we obtain a trimer binding energy  significantly
smaller,  $-1.9(4)$ mK. The trimer energy is therefore sizably affected by
the interatomic potential and the reason for that lies on the huge
cancellation between the kinetic and potential energies. For instance,
using the JDW potential and the pure estimator for the potential energy, we
obtain $E_p=-355(9)$ mK and $E_k=350(9)$ mK. For larger clusters, the
influence of the potential form on the binding strength is reduced: for the
tetramer and with the Silvera potential,  we obtain an energy $E =-107(2)$
mK  which is only about 15\% weaker than the one for the JDW potential,
$E=-126(2)$ mK.   The potential and kinetic energies of the tetramer using
the JDW potential are $E_p=-1871(11)$ mK and $E_k=1745(12)$ mK.

The addition of one H$\downarrow$ atom to the (T$\downarrow)_N$ cluster
creates a system that appears to be at the threshold of stability or
unstable, that is the ground-state energy of the
(T$\downarrow)_N$H$\downarrow$  cluster is within the errorbars equal to or
higher than the energy of the (T$\downarrow)_N$ cluster, respectively. In
order to be certain about the conclusion concerning the stability of the 
(T$\downarrow)_N$H$\downarrow$ clusters, we have repeated the calculations
using two different types of trial wave functions. In the case where VMC
energies are further away from the DMC ones, a lot of simulation time is
needed for the system to reach the equilibrium state, and in some cases
with the trial function (\ref{trial2}) the lowering of the energy with
imaginary time going on is almost imperceptible.  On the other hand, with
both (\ref{trial1}) and (\ref{trial3}), and for small values of the
parameter $s$ or $b_3$ in the $f(r_{TH})$ function, the system rather
quickly reaches the equilibrium energy, which is within the errorbars equal
to the energy of the system without the H$\downarrow$ atom. Furthermore, if
we continue the simulation after the system has reached the ground-state
energy, we can observe the surplus H$\downarrow$ atom leaving the cluster.
This indicates that these clusters are effectively unstable. We have also
calculated the ground-state energies of several clusters with more hydrogen
atoms (T$\downarrow)_N$(H$\downarrow)_M$, such that $N+M \le 10$. Our
results indicate that these clusters are also unstable, because although in
some cases negative energies are obtained they are  above those for
(T$\downarrow)_N$H$\downarrow$.

\begin{figure}[p]
\centering
        \includegraphics[width=8.5cm]{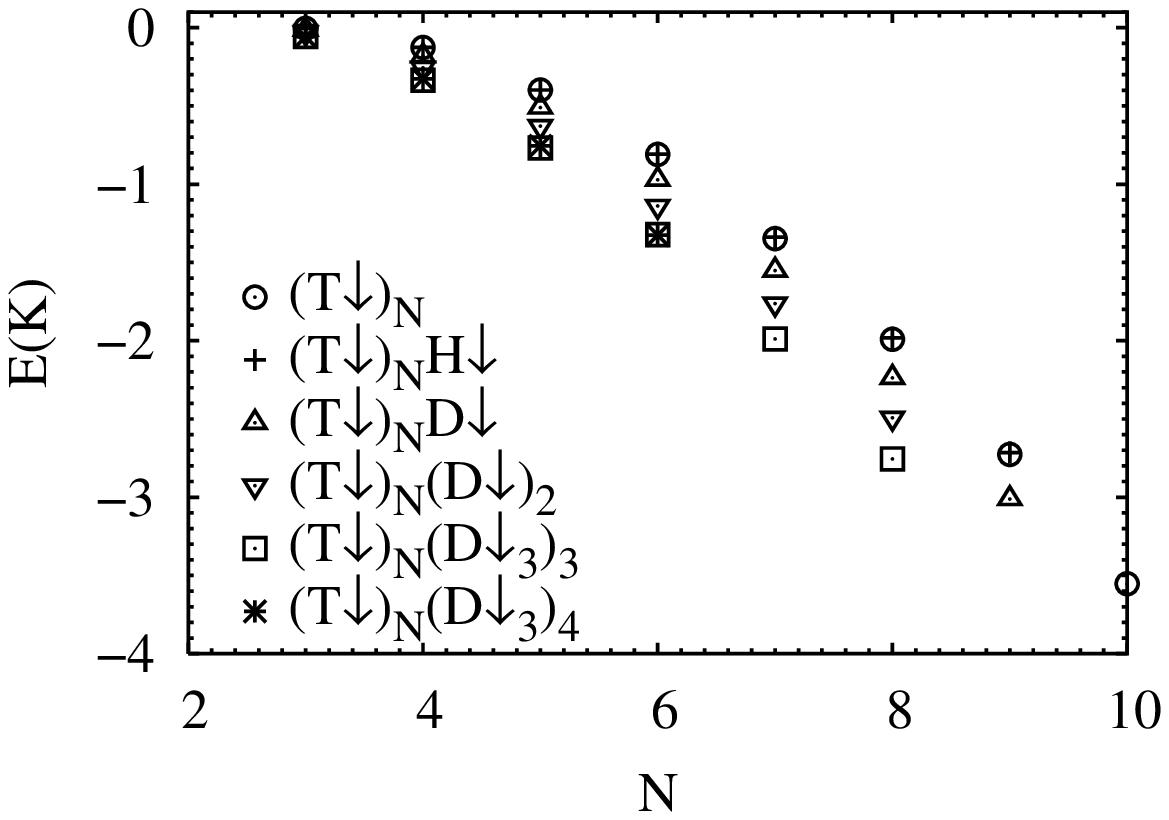}
        \caption{I. Be\v{s}li{\'c} {\it et al.}} 
	\label{fig:allen}
\end{figure}

 Table \ref{tab:energiesTD} presents the results for the ground-state
 energy of  mixed T$\downarrow$-D$\downarrow$ clusters. Due to the
 approximately twice larger mass of D$\downarrow$ compared to
 H$\downarrow$,  clusters with one or two  D$\downarrow$ atoms and at least
 three T$\downarrow$ atoms are stable. In the clusters
 (T$\downarrow)_N$(D$\downarrow)_2$,  we have assumed that D$\downarrow$ 
 atoms occupy two different nuclear spin states. For clusters
 (T$\downarrow)_N$(D$\downarrow)_M$ with $M>$2, we have considered  both
 the case where two different nuclear spin states are occupied
 (T$\downarrow)_N$(D$\downarrow_2)_M$ and the case with three different
 occupied nuclear spin states (T$\downarrow)_N$(D$\downarrow_3)_M$. Similar
 to the case of the bulk system,~\cite{panoff,flynn,skjetne}  we find that
 the three-component spin clusters are more strongly bound than the ones
 where D$\downarrow$ occupies two nuclear spin states,
 $E(($T$\downarrow)_N($D$\downarrow_3)_M) <
 E(($T$\downarrow)_N($D$\downarrow_2)_M)$. Furthermore,
 (T$\downarrow)_N$(D$\downarrow_3)_3$, whose spatial wave function has no
 nodes, are all stable for $N \ge 3$, while it takes  6 T$\downarrow$ atoms
 to form a stable (T$\downarrow)_N$(D$\downarrow_2)_3$ cluster. The smaller
 (T$\downarrow)_N$(D$\downarrow_2)_3$  clusters have the same energy as the
 (T$\downarrow)_N$(D$\downarrow)_2$ clusters within the errorbars,  but the
 analysis of the distributions shows one of the D$\downarrow$ atoms leaving
 the system. The clusters (T$\downarrow)_N$(D$\downarrow_3)_4$ are
 similarly at the threshold of binding for $N=3,4,5,6$, just like the
 clusters (T$\downarrow)_N$(D$\downarrow_2)_4$, and again it is necessary
 to consider the structure to better determine their stability.  Our
 analysis indicates that only (T$\downarrow)_6$(D$\downarrow_3)_4$ is
 stable. The same behavior, as far as the energy is concerned, is
 reproduced for (T$\downarrow)_N$(D$\downarrow_3)_5$ with  $N\le 5$: as the
 separation of the D$\downarrow$ atoms is growing along the simulation we
 are led to consider them unstable. 
 We expect that all other clusters
 (T$\downarrow)_N$(D$\downarrow)_M$, with $N+M\le 10$ and $M>5$ are unstable
 because the exchange of T$\downarrow$ with the D$\downarrow$ atoms does
 not  change the interaction but raises the kinetic energy.   
 The evolution of the energy of  mixed clusters formed by   $N$ 
 tritium atoms and one hydrogen atom or up to four deuterium
 atoms is shown in Fig. \ref{fig:allen}.

In addition to the energy, DMC simulations allow also for exact estimations
of other relevant magnitudes such as the distribution of interparticle
distances $P(r)$ or the distribution of particles with  respect to
the center of mass of the cluster $\rho(r)$. In both cases  
it is possible to eliminate the bias
coming from the trial wave function by using pure estimators~\cite{pures}
 and arrive to exact results. 

\begin{figure}[p]
\centering
        \includegraphics[width=8.5cm]{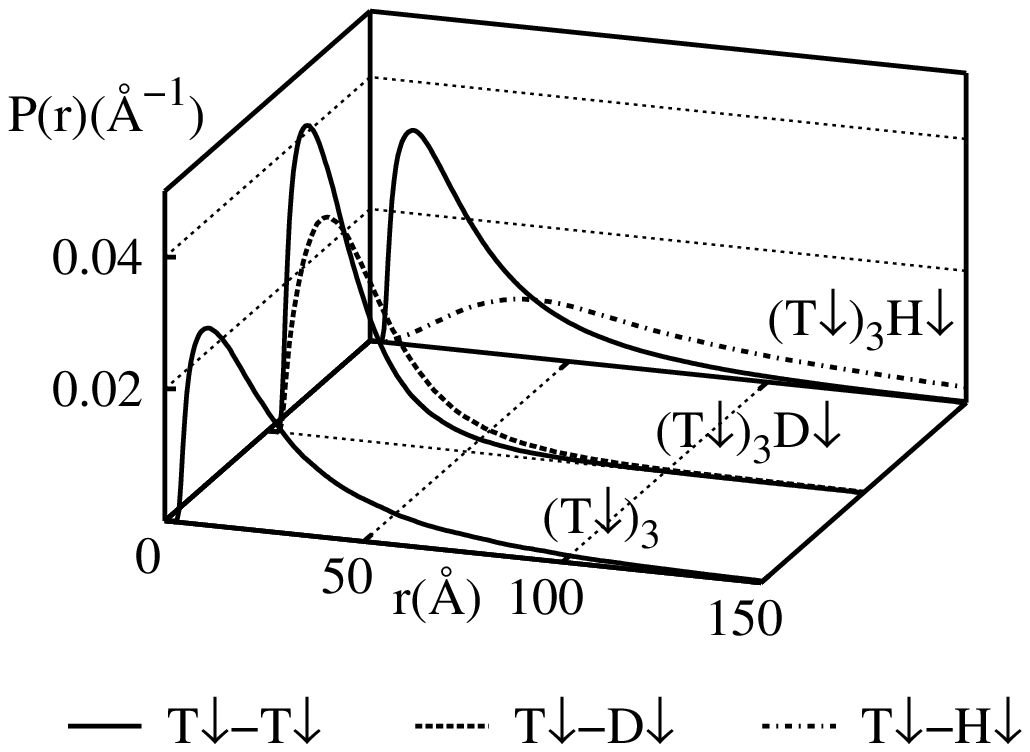}%
        \caption{ I. Be\v{s}li{\'c} {\it et al.} }
	\label{fig:pair3T1D1H}
\end{figure}

Our calculations confirm that the pure T$\downarrow$ clusters are spatially
very diffuse.~\cite{blume} From the DMC results for $P(r)$ one can see that
the average separation between particles $\langle r_{TT} \rangle$ ranges
between  34 \AA\, in the case of (T$\downarrow)_3$ to 10.9 \AA\, for
(T$\downarrow)_{6}$ and 10.5 \AA\, for (T$\downarrow)_{10}$. The addition
of one D$\downarrow$ or one H$\downarrow$ reduces the average separation
between T$\downarrow$ atoms in the case of clusters with $N=3$ or 4, as can
be seen in  Fig. \ref{fig:pair3T1D1H}, which compares the pair distribution
functions of  (T$\downarrow)_3$, (T$\downarrow)_3$D$\downarrow$ and
(T$\downarrow)_3$H$\downarrow$. In contrast, for clusters with more
T$\downarrow$ atoms the T$\downarrow$-T$\downarrow$ separation remains
almost unaffected by the addition of one D$\downarrow$ or one H$\downarrow$
atom. Due to the larger zero-point motion of D$\downarrow$ and
H$\downarrow$ atoms,  $\langle r_{TT} \rangle ~>~  \langle r_{TD} \rangle
~>~ \langle r_{TH} \rangle$. Moreover, we obtain $\langle r_{TH} \rangle$
greater than 100 \AA\, indicating that the cluster
(T$\downarrow)_3$H$\downarrow$ is unstable. Similar behavior is noticed for
larger (T$\downarrow)_N$H$\downarrow$ clusters.

\begin{figure}[p]
\centering
        \includegraphics[width=8.5cm]{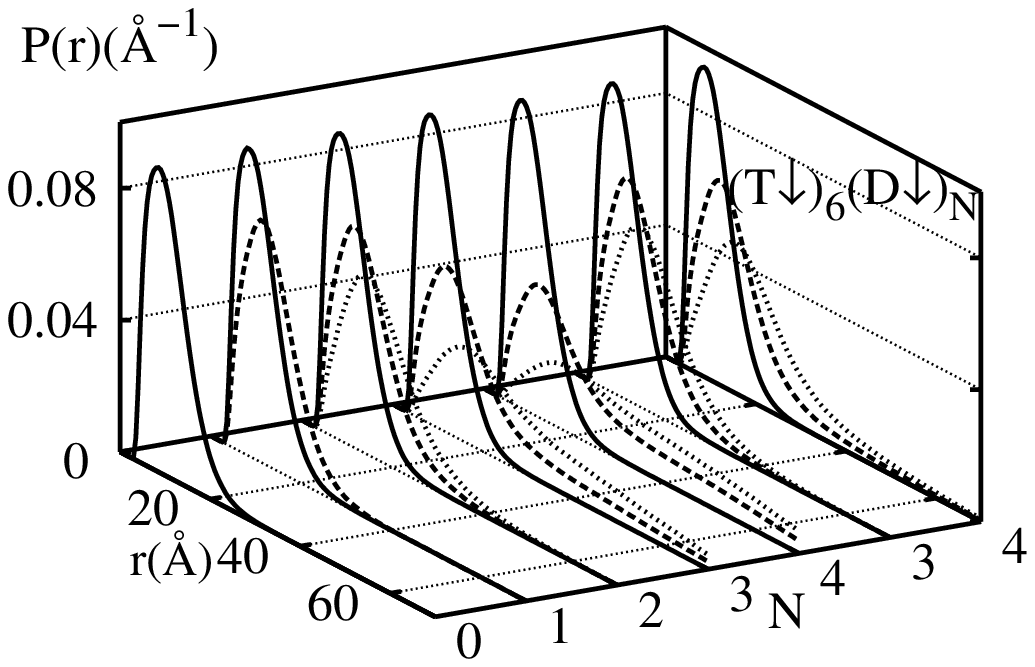}%
        \caption{ I. Be\v{s}li{\'c} {\it et al.}}
	\label{fig:pair7TnD}
\end{figure}

The effect of adding D$\downarrow$ atoms to the core of (T$\downarrow)_6$
cluster is shown in Fig. \ref{fig:pair7TnD}. The separation between pairs
of T$\downarrow$ atoms remains almost unchanged, while the
T$\downarrow$-D$\downarrow$ and D$\downarrow$-D$\downarrow$ separations
grow with the addition of D$\downarrow$ atoms. There is a noticeable
difference between clusters with D$\downarrow_2$ and  D$\downarrow_3$
spin-polarized deuterium. Former clusters with more than three particles 
are much more extended, i.e., the average separation between D$\downarrow$
atoms is larger and the tail of the distribution decays slowly.  Despite
the large size of the cluster, the average separation between particles
does not appear to grow in the case of (T$\downarrow)_6$(D$\downarrow_2)_3$
so we are led to consider this cluster as stable; the same conclusion 
applies to    (T$\downarrow)_6$(D$\downarrow_3)_{4}$. On the other hand,
the average  D$\downarrow$-D$\downarrow$ distance in the
(T$\downarrow)_6$(D$\downarrow_2)_{4}$ cluster is slightly above 40 \AA\,
but grows very slowly in the course of the simulation, indicating that it
could be unstable.

\begin{figure}[p]
\centering
        \includegraphics[width=8.5cm]{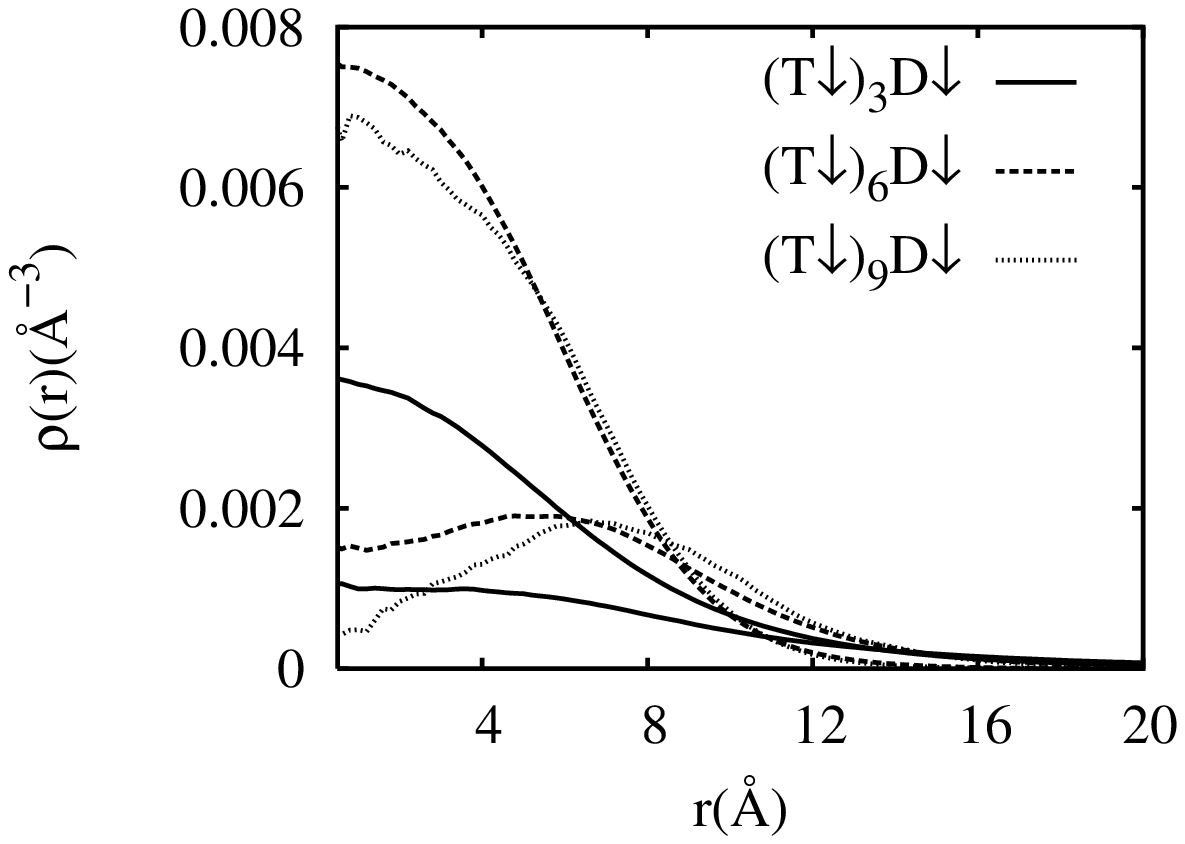}%
        \caption{I. Be\v{s}li{\'c} {\it et al.}  }
	\label{fig:TnD}
\end{figure}

Finally, Fig. \ref{fig:TnD} presents the density distribution of the clusters
(T$\downarrow)_N$(D$\downarrow)$, for $N=$3,6,9. With the increase in the
number of bosons, deuterium is pushed to the  surface of the cluster,
although even for $N=9$ there is still an appreciable probability of finding it
inside the cluster. Similar behavior is found for
(T$\downarrow)_N$(H$\downarrow)$ clusters. For larger clusters, one can expect
that both D$\downarrow$  and H$\downarrow$ will be pushed to the surface
forming the so-called Andreev states,~\cite{Andreev} similarly to the
well known behavior of $^3$He in mixed $^3$He-$^4$He clusters.~\cite{review}

\section{conclusions}

The ground-state properties of spin-polarized pure and mixed tritium
clusters have been accurately determined using the DMC method.  Our results
show that  the trimer is a Borromean or halo state in agreement with the results
of Ref. \onlinecite{blume}. The most promising
candidates for the super-Borromean states in mixed clusters have been
(T$\downarrow)_2$(D$\downarrow)_{1,2,3,4,5}$, but after a careful analysis
we have found no bound
states in either of these clusters. On the other hand, we conclude that 
it takes at least three T$\downarrow$
atoms in order to bind 1-3 D$\downarrow$ atoms, the case of 
(T$\downarrow)_3$(D$\downarrow)_3$ being bound only when the deuterium atoms are
of D$\downarrow_3$ type. The DMC results show that 6 T$\downarrow$ atoms are
needed in order to bind 4  D$\downarrow_3$ atoms, while
clusters with up to total 10 atoms and more than 4 D$\downarrow$ atoms 
seem all to be unstable.    We have not considered mixed clusters with the
D$\downarrow$ atoms in only one nuclear spin-state D$\downarrow_1$ 
since,  from analogy with the bulk, we expect them to be less bound than
D$\downarrow_2$ and D$\downarrow_3$. 

Finally, and concerning mixed tritium-hydrogen clusters,
our results show that all the (T$\downarrow)_N$(H$\downarrow)_M$ clusters with $N+M \le 11$ 
are unstable. Work is in progress to find out how many 
T$\downarrow$ atoms are needed to bind one H$\downarrow$ atom.

\acknowledgments
J. B. acknowledges support from DGI (Spain) Grant No. FIS2005-04181 and
Generalitat de Catalunya Grant No. 2005SGR-00779. I.B and L.V.M. acknowledge
support from MSES (Croatia) Grant No. 177-1770508-0493. We also acknowledge
the support of the Central Computing Services at the Johannes Kepler
University in Linz, where part of the computations was performed.


\begin{table}[p]
\begin{center}
\begin{tabular}{c|cc}       
~~N~~& ~~~~(T$\downarrow)_N$~~ &  ~~~~(T$\downarrow)_N$H$\downarrow$~~ 
\\ \hline
2 &  -  &  - \\
3  & -4.8 (0.2) &\it{-4.7(0.7)}  \\
4 &-126(2) & \it{-126(1)} \\
5  & -398(1) & \it{-398(2)}\\
6 & -810(2) & \it{-807(3)}\\ 
7  & -1348(4) &\it{-1339(6)}\\
8 & -1991(5) &  \it{-1982(7)} \\
9 & -2727(7) &  \it{-2720(9)}\\
10 &-3553(8) & 
\end{tabular} 
\end{center}
\caption{The ground state energy (in mK) of  pure spin-polarized tritium  
clusters with $N$ atoms and tritium clusters with an additional hydrogen
atom. Clusters which appear to be unstable are written in italic. 
Figures in parenthesis are the statistical errors.} 
\label{tab:energiesTH}
\end{table}


\begin{table}[p]
\begin{center}
\begin{ruledtabular} 
\begin{tabular}{c|ccccccc} 
  \multicolumn{8}{c}{E[(T$\downarrow)_N$(D$\downarrow)_M$]} \\ \hline   
N&  D$\downarrow$ & (D$\downarrow)_2$ & (D$\downarrow_2)_3$ & 
(D$\downarrow_2)_4$ & (D$\downarrow_3)_3$ & (D$\downarrow_3)_4$ & 
(D$\downarrow_3)_5$\\ \hline
2 &  -  & - &  - & -& -  & -& -\\
3   & -12.2(0.9) & -31(3)& \it{-27(3)}&\it{-21(2)}&-56(2) & \it{-58(5)} & \it{-46(6)} \\
4  & -182 (3)& -256(3)& \it{-253(4)}&\it{-241(3)}& -336(3) &\it{-327(6)} & \it{-313(6)}\\
5   & -510 (4)&-629(4) &\it{-633(6)} &\it{-610(6)}&-769(4) &\it{-755(7)} & \it{-732(6)}\\
6  &-972(4) & -1139(3)& \it{-1145(9)}&\it{-1135(5)}& -1322(5)&\it{-1326(10)} &\\ 
7   &-1552(4) & -1763(10 )&\it{-1747(15)}& &-1989(7)& &\\\
8  & -2237(6)&-2492(12) & -2554(15)&& -2755(8) & &\\
9  & -3013(7)& & & & &
\end{tabular} 
\end{ruledtabular}
\end{center}
\caption{The ground-state energy $E$ (in mK) of  mixed spin-polarized 
tritium-deuterium clusters.
For mixed clusters having more than two deuterium atoms, 
combinations with D$\downarrow_2$  and D$\downarrow_3$ are considered.
The clusters who are at the threshold of binding or appear to be 
unstable are written in italic. Figures in parenthesis are the statistical errors.} 
\label{tab:energiesTD}
\end{table}


\textbf{Figure captions}\\

FIG. 1: Comparison of our results for spin-polarized tritium 
	clusters (circles) with the results of Blume \textit{et. al.}~\cite{blume} (crosses). 
	The difference between our results and the ones from Blume \textit{et. al.}~\cite{blume}
	is mainly due to the use of a slightly different interaction (see text). 
	The error bars of the DMC energies are smaller than the size of the 
	symbols.

FIG. 2: Energies of the spin-polarized pure and mixed 
	tritium clusters in K as a function of the number of 
	T$\downarrow$ atoms.
	
FIG. 3: Distribution of interparticle distances in the 
	(T$\downarrow)_3$, (T$\downarrow)_3$D$\downarrow$ and 
	(T$\downarrow)_3$H$\downarrow$ clusters.

FIG. 4: Distribution of interparticle distances of the 
	(T$\downarrow)_6$(D$\downarrow)_N$ clusters. Full line 
        denotes T$\downarrow$-T$\downarrow$, dashed line 
	T$\downarrow$-D$\downarrow$, and dotted line 
	D$\downarrow$-D$\downarrow$ separation. 
	$N=3, 4$ distributions with lower maxima in 
	D$\downarrow$-D$\downarrow$ separations correspond to  
	D$\downarrow_2$ and those with higher maxima to  
	D$\downarrow_3$.
	
FIG. 5: Density distributions of T$\downarrow$ (upper curves) 
	and D$\downarrow$ (lower curves) 
	for (T$\downarrow)_N$D$\downarrow$ clusters.

\end{document}